\documentclass{kapproc} % Computer Modern font calls

\usepackage{procps} 

\usepackage[dvips]{graphicx}

\upperandlowercase

 \let\footnote\savefootnote

\normallatexbib %will produce bibliography entries as shown in the
\begin{document}

\articletitle {The Turbulent Interstellar Medium: Insights and
Questions from Numerical Models
}

\chaptitlerunninghead{Turbulent ISM} % Shorter running head title.

\author{Mordecai-Mark Mac Low,\altaffilmark{1} Miguel A. de
Avillez,\altaffilmark{2} \& Maarit J. Korpi\altaffilmark{3,4}}

\affil{
\altaffilmark{1}Department of Astrophysics, American Museum of
Natural History, 79th Street at Central Park West, New York, NY,
10024-5192, USA, mordecai@amnh.org \\ 
\altaffilmark{2}Department of Mathematics, University of \'Evora,
R. Rom\~ao Ramalho 59, P-7000 \'Evora, Portugal,
mavillez@galaxy.lca.uevora.pt \\
\altaffilmark{3} Division of Astronomy, University of Oulu, PO Box
3000, FIN-90014 University of Oulu, Finland, Maarit.Korpi@oulu.fi \\
\altaffilmark{4} Laboratoire d'Astrophysique, Observatoire
Midi-Pyr\'en\'ees, 14 Avenue Edouard Belin, F-31400 Toulouse, France }

\begin{abstract}

{\em "The purpose of numerical models is not numbers but insight."}
(Hamming) In the spirit of this adage, and of Don Cox's approach to
scientific speaking, we discuss the questions that the latest
generation of numerical models of the interstellar medium raise, at
least for us.  The energy source for the interstellar turbulence is
still under discussion.  We review the argument for supernovae
dominating in star forming regions.  Magnetorotational instability has
been suggested as a way of coupling disk shear to the turbulent flow.
Models make evident that the unstable wavelengths are very long
compared to thermally unstable wavelengths, with implications for star
formation in the outer galaxy and low surface brightness disks. The
perennial question of the factors determining the hot gas filling
factor in a SN-driven medium remains open, in particular because of
the unexpectedly strong turbulent mixing at the boundaries of hot
cavities seen in the models.  The formation of molecular clouds in the
turbulent flow is also poorly understood.  Dense regions suitable for
cloud formation clearly form even in the absence of self-gravity,
although their ultimate evolution remains to be computed.

\end{abstract}

\section{Questions about Turbulence}
Numerical models often yield insight into the behavior of a physical
system long before they can give quantitative results.  In this
contribution we review possible answers to three major questions about
turbulence, relying on a combination of general energetic arguments
and numerical models.  

The first question is, ``What provides the energy to drive the turbulent
flow?''  Many sources have been proposed, but few have the required
energy to counteract dissipation in the interstellar medium.
Supernovae (SNe) seem likely to be the primary driver in parts of
galaxies where star formation occurs, while the magnetorotational
instability (MRI) may couple the gas to galactic rotational shear in other
parts of galaxies. 

The second question is, ``How does the driving shape the flow?'' Most
of the energy lies at the driving scale, so the large-scale structure
is determined quite directly by the driving mechanism.  Turbulent
compression may be as important as thermodynamic phases in determining
the pressure at any particular point in the ISM, as well as in
determining the filling factor of the hot gas.

The last question, of interest to understanding the rate of star
formation from the ISM, is ``How do molecular clouds form in this
flow?'' Turbulent compression and self-gravity both appear as possible
mechanisms, but cannot yet be definitively distinguished.

\section{What Drives the Turbulence?}
Maintenance of observed motions appears to depend on continued driving
of the turbulence, which has kinetic energy density $e = (1/2) \rho
v_{\rm rms}^2$.  Mac Low \cite{m99,m02} estimates that the dissipation
rate for isothermal, supersonic turbulence is
\begin{eqnarray} \label{eqn:dissip}
\dot{e}  & \simeq & -(1/2)\rho v_{\rm rms}^3/L_{\rm d} \\
 & = & -(3 \times 10^{-27} \,\mbox{erg}\,\mbox{cm}^{-3}\,\mbox{s}^{-1}) 
\left(\frac{n}{1\,\mbox{cm}^{-3}}\right)
\left(\frac{v_{\rm rms}}{10\,\mbox{km}\,\mbox{s}^{-1}}\right)^3 
\left(\frac{L_{\rm d}}{100 \,\mbox{pc}}\right)^{-1}, \nonumber
\end{eqnarray}
where $L_{\rm d}$ is the driving scale, which we have somewhat
arbitrarily taken to be 100$\,$pc (though it could well be smaller, or
a broad range), and we have assumed a mean mass per particle $\mu=
2.11\times 10^{-24}$~g.  The dissipation time for turbulent kinetic
energy
%\begin{equation} \label{eqn:disstime}
$\tau_{\rm d} = e / \dot{e} \simeq L_{\rm d}/v_{\rm rms},$
% = (9.8\, \mbox{Myr})
%\left(\frac{L_{\rm d}}{100\,\mbox{pc}}\right)
%\left(\frac{v_{\rm rms}}{10\,\mbox{km}\,\mbox{s}^{-1}}\right)^{-1},
%\end{equation}
which is just the crossing time for the turbulent flow across the
driving scale \cite{e00}.  What then is the energy source for
this driving? We here review the energy input rates for the most
plausible mechanisms, feedback from massive stars, particularly SNe,
and magnetorotational instabilities.  A more extensive discussion
covering a number of other possibilities as well is given by Mac Low
\& Klessen \cite{mk04}.

An energy source for interstellar turbulence that has long been
considered is shear from galactic rotation \cite{f81}.  However, the
question of how to couple from the large scales of galactic rotation
to smaller scales remained open.  Sellwood \& Balbus \cite{sb99} suggested
that the MRI \cite{bh91,bh98} could couple the large and
small scales efficiently.  The instability generates Maxwell stresses
(a positive correlation between radial $B_R$ and azimuthal $B_{\Phi}$
magnetic field components) that transfer energy from shear into
turbulent motions at a rate
\begin{equation}
\label{eqn:stress}
\dot{e} = - T_{R\Phi} (d\Omega / d \ln R) =  T_{R\Phi} \Omega,
\end{equation}
where the last equality holds for a flat rotation curve \cite{sb99}.
Numerical models suggest that the Maxwell stress tensor
$T_{R\Phi} \simeq 0.6 B^2/(8\pi)$ \cite{h96}.
For the Milky Way, the standard value of the rotation rate $\Omega =
(220 \mbox{ Myr})^{-1} = 1.4 \times 10^{-16} \mbox{ rad s}^{-1}$, so
the MRI can contribute energy at a rate
\begin{equation}
\dot{e} = (3 \times 10^{-29}\,\mbox{erg}\,\mbox{cm}^{-3}\,\mbox{s}^{-1})
\left(\frac{B}{3 \mu\mbox{G}}\right)^2 \left(\frac{\Omega}{(220\,\mbox{Myr})^{-1}}\right). 
\end{equation}
For parameters appropriate to the H{\sc i} disk of a sample small
galaxy, NGC~1058, including $\rho = 10^{-24}\,$g$\,$cm$^{-3}$,
Sellwood \& Balbus \cite{sb99} find that the magnetic field required to
produce the observed velocity dispersion of 6~km~s$^{-1}$ is roughly 3
$\mu$G, a reasonable value for such a galaxy.  This instability may
provide a base value for the velocity dispersion below which no galaxy
will fall.  If that is sufficient to prevent collapse, little or no
star formation will occur, producing something like a low surface
brightness galaxy with large amounts of H{\sc i} and few stars. This
may also apply to the outer disk of our own Milky Way and other
star-forming galaxies.

In active star-forming galaxies, massive stars probably dominate the
driving at radii where they form.  They could do so through ionizing
radiation and stellar winds from O~stars, or clustered and field
SN explosions, predominantly from B~stars no longer associated
with their parent gas.  Mac Low \& Klessen \cite{mk04} demonstrate
that ionizing radiation is unlikely to dominate the kinetic energy
budget, despite the large amount of energy going into heating and
ionization. The total energy
input from the line-driven stellar wind over the main-sequence lifetime
of an early O~star can equal the energy from its SN explosion,
and the Wolf-Rayet wind can be even more powerful.  However, the
mass-loss rate from stellar winds drops as roughly the sixth power of
the star's luminosity,
% if we take into account that stellar luminosity
%varies as the fourth power of stellar mass (Vink, de Koter \& Lamers
%2000), 
while the powerful Wolf-Rayet winds \cite{nl00} last
only $10^5$ years or so, so only the very most massive stars
contribute substantial energy from stellar winds.  The energy from
SN explosions, on the other hand, remains nearly constant down
to the least massive star that can explode.  As there are far more
lower-mass stars than massive stars, 
%with a Salpeter IMF giving a
%power-law in mass of $\alpha = -2.35$,
%(Eq.\ \ref{eqn:salpeter}), 
SN explosions inevitably dominate over stellar winds after the
first few million years of the lifetime of an OB association.

To estimate the energy input rate from SNe,
%we begin by finding the SN rate in the Galaxy $\sigma_{SN}$.
%Cappellaro et al.\ (1999) estimate the total SN rate in
%SN units to be $0.72 \pm 0.21$ SNu for galaxies of type S0a-b
%and $1.21 \pm 0.37$~SNu for galaxies of type Sbc-d, where 1 SNu = 1 SN
%(100~yr)$^{-1} (10^{10} L_B/\mbox{L}_{\odot})^{-1}$, and $L_B$ is the
%blue luminosity of the galaxy.  Taking the Milky Way as lying between
%Sb and Sbc, we estimate $\sigma_{SN} = 1$~SNu.  Using a Galactic
%luminosity of $L_B = 2 \times 10^{10} \mbox{ L}_{\odot}$, we find a
we begin with a SN rate for the Milky Way of (50~yr)$^{-1}$, which
agrees well with the estimate in equation~(A4) of McKee \cite{m89}).  If we
take the scale height of SNe $H_c \simeq 100$~pc and a star-forming
radius $R_{sf} \simeq 15$~kpc, we can compute the energy input rate from SN
explosions with energy $E_{SN} = 10^{51}\,$erg to be
\begin{eqnarray}
\dot{e} & = &\frac{\sigma_{SN} \eta_{SN} E_{SN}}{\pi R_{sf}^2 H_c} \\
       &  = & (3 \times 10^{-26} \mbox{ erg s$^{-1}$ cm}^{-3})
\left(\frac{\eta_{SN}}{0.1} \right)
\left(\frac{\sigma_{SN}}{1 \mbox{ SNu}} \right) 
\left(\frac{H_c}{100 \mbox{ pc}} \right)^{-1} \times \nonumber \\
& \times & \left(\frac{R_{sf}}{15 \mbox{ kpc}} \right)^{-2}
\left(\frac{E_{SN}}{10^{51} \mbox{ erg}} \right). \nonumber
\end{eqnarray}
The efficiency of energy transfer from SN blast waves to the
interstellar gas $\eta_{SN}$ depends on the strength of radiative
cooling in the initial shock, which will be much stronger in the
absence of a surrounding superbubble (e.g.\ \cite{h90}).  Substantial
amounts of energy can escape in the vertical direction in superbubbles
as well, however.  
%Norman \& Ferrara (1996) make an analytic estimate
%of the effectiveness of driving by SN remnants and superbubbles. 
The scaling factor $\eta_{SN} \simeq 0.1$ used here was derived by
Thornton et al.\ \cite{t98} from detailed, 1D, numerical simulations of
SNe expanding in a uniform ISM.  It can alternatively be drawn from
momentum conservation arguments,
%(eq.~\ref{eqn:mntm-cons}), 
comparing a typical expansion velocity of 100$\,$km$\,$s$^{-1}$ to
typical interstellar turbulence velocity of 10$\,$km$\,$s$^{-1}$.
Multi-dimensional models of the interactions of multiple SN remnants
(e.g.\ \cite{a00}) are required to better determine the effective
scaling factor.

SN driving appears to be powerful enough to maintain the
turbulence even with the dissipation rates estimated in
Eq.~(\ref{eqn:dissip}).  It provides a large-scale
self-regulation mechanism for star formation in disks with sufficient
gas density to collapse despite the velocity dispersion produced by
the MRI.  As star formation increases in
such galaxies, the number of OB stars increases, ultimately increasing
the SN rate and thus the velocity dispersion, which restrains
further star formation.

\section{How Does Driving Shape the ISM?}

We now turn to the question of how these different driving mechanisms
determine the structure of the ISM.  Clearly, different mechanisms
yield different results.

To study the MRI, we used a parallel MHD code integrating $\ln \rho$
rather than density $\rho$ to handle strong density contrasts
\cite{ck01}, with shearing sheet horizontal boundary conditions
implemented.  The preliminary models shown here were run at $64 \times
64 \times 128$ zones on an $0.5 \times 0.5 \times 1$~kpc grid, with
the ISM in vertical hydrostatic equilibrium with scale height $H =
250$~pc initially, and an initially vertical magnetic field with
thermal to magnetic pressure ratio $\beta = 1000$.  The initial
wavelength of maximum instability was then 80~pc.  Runs were extended
to 10 orbits, or 2.5~Gyr.  Radiative cooling was included based on an
equilibrium ionization cooling curve including thermal instability
below $10^4$~K, and heating proportional to density was chosen to
exactly balance the cooling in the initial model.  We ran models
initially in thermally stable and unstable regimes.

\begin{figure}[thb]
\centerline{
\includegraphics[width=\textwidth,angle=270]{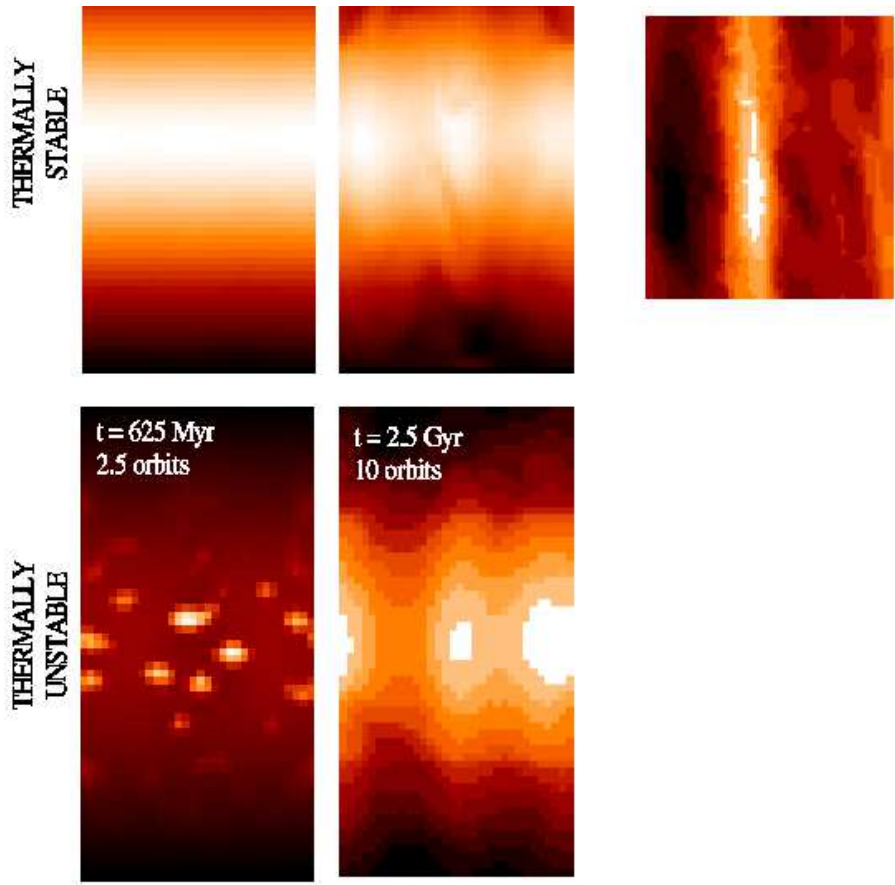}
}
%\vskip -1.8in
\caption{\label{mri}
Log of density on vertical cuts through 3D shearing sheet models of
MRI at times given. {\em Top:} thermally stable models, including a
horizontal cut through the midplane of the final model; {\em Bottom:}
thermally unstable models, showing the action of the MRI on the cooled
clumps.}
\end{figure}
In Figure~\ref{mri} we show the development of the MRI in these
regimes.  In the thermally stable regime, factor of 2--3 column
density contrasts through the disk are created by the instability.  In
the thermally unstable regime, the thermal instability acts quickly to
clump the gas, but after multiple orbits the MRI adds sufficient
velocity dispersion to heat the gas and distribute it more uniformly.
Rather more substantial column density contrasts still occur.
Comparison with observed H{\sc i} disks outside of the star-forming
region should be revealing of whether this mechanism is in fact
maintaining their velocity dispersion.

Numerical models of the SN-driven ISM suggest that the hot gas filling
factor $f$ is closer to the value $f \sim 0.2$ \cite{a00,a04,ab03} predicted
by Slavin \& Cox \cite{sc93} than to the values close to unity predicted by
McKee \& Ostriker \cite{mo77}.  Why is this?  McKee \& Ostriker \cite{mo77}
assumed a two-phase medium with cold, dense clouds embedded in a
uniform density, warm, intercloud medium.  Hot SN remnants then
expanded into this medium.  Was the cooling within the SN remnants
underestimated because turbulent mixing was approximated with mass
loading from the clumps overrun by the remnants, or was the effective
external density underestimated by the two-phase model?

\begin{figure}[thb]
\centerline{
\includegraphics[width=0.5\textwidth,angle=270]{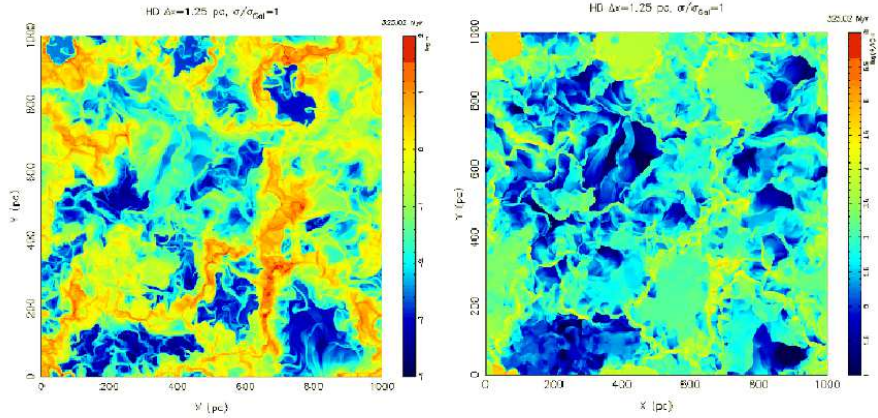}
}
%\vskip -2.1in
\caption{\label{sne} Cuts through the midplane of the SN-driven model
run at finest resolution of 1.25 pc.  {\em Left:} Density; {\em
Right:} Pressure.}
\end{figure}
To study the SN-driven ISM we used an adaptive mesh refinement
code described by Avillez \& Mac Low \cite{am02}, with a $1 \times 1
\times 20$~kpc grid set up as described in Avillez \cite{a00}, with a SN
rate equal to the galactic value. 

\begin{figure}[thb]
%\vskip -0.2in
\centerline{
\includegraphics[width=0.6\textwidth,angle=270]{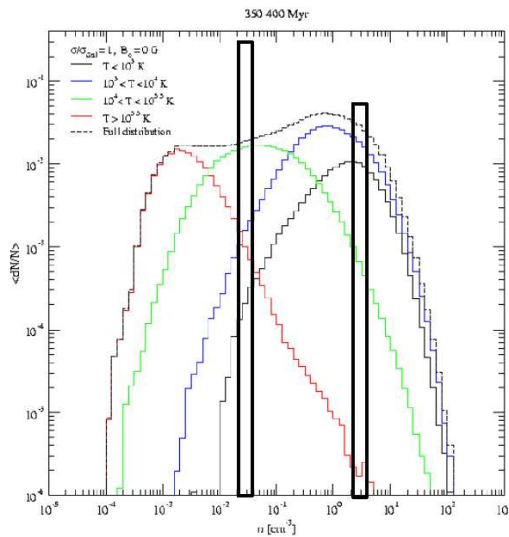}
}
%\vskip -0.2in
\caption{\label{rho-pdf}
Probability distribution function of density in the midplane of the
SN-driven model.  The two-phase medium assumed by McKee \& Ostriker
\cite{mo77} is shown schematically by the thick black rectangles. }
\end{figure}

Figure~\ref{sne} shows that the pressures vary widely \cite{m04}, so that no
simple two-phase medium can actually form.  Instead, the densities
cover a broad range continuously, as shown in Figures~\ref{sne}
and~\ref{rho-pdf}.
This continuous distribution of density may act to impede the
expansion of SN remnants more effectively than the warm intercloud
medium with cold embedded clouds shown schematically in
Figure~\ref{rho-pdf}.  

On the other hand, Avillez \& Mac Low \cite{am02} demonstrated using a
tracer field that mixing occurs quite efficiently in the hot regions.
In Figure~\ref{sne} widespread turbulent mixing at the edges of shells
and supershells can be seen.  This could substantially enhance the
density in the hot interiors, thus enhancing the radiative cooling,
which is proportional to $\rho^2$.  However, a quantitative test of
how well or poorly this turbulent mixing was modeled by the model of
SN remnants overrunning conductively evaporating clouds used by McKee
\& Ostriker \cite{mo77} remains to be done.

\section{How Do Molecular Clouds Form in the Turbulent ISM?}

\begin{figure}[thb]
%\vskip -0.4in
%\centerline{
\includegraphics[width=0.5\textwidth,angle=270]{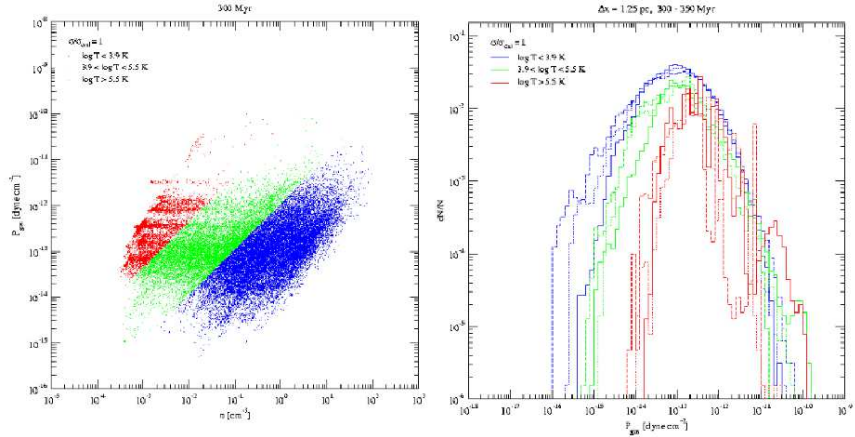}
%}
%\vskip -1.8in
\caption{\label{press} {\em Left:} Scatter plot of pressure vs.\
density in the midplane of the SN-driven model showing occupation of
high-pressure, high-density region associated with molecular clouds.
{\em Right:} Volume-weighted probability distribution function of
pressure in the same model. Note that the small volumes occupied by
high-density, cold gas have large mass. }
\end{figure}

Molecular clouds are high-density objects, with much of their mass at
densities of $10^{3-5}$~cm$^{-3}$.  With typical temperatures of order
10~K, their pressures are an order of magnitude or more above the
average ISM pressure.  It has usually been argued that these high
pressures must be caused by self-gravity, since they would otherwise
explode.  However, turbulent ram pressure in a SN-driven ISM produces
high-density, high-pressure regions even in the absence of
self-gravity, as shown in Figure~\ref{press}.  These may provide the
sites for the formation of at least some molecular clouds, especially
ones that do not show vigorous, efficient, star formation.

\begin{acknowledgments}
M-MML is partly supported by NSF grants AST 99-85392 and AST 03-07854.
This work made use of the NASA ADS Abstract Service.
\end{acknowledgments}

\section{Discussion}

\noindent
{\it Gaensler:} We know from observations of scattering \&
scintillation that there is turbulence in the warm ionized phase of
the ISM. Since we expect expanding SN remnants to sweep up neutral
shells, can SN remnants still drive the turbulence seen in ionized
gas?  What sort of a contribution do ionization fronts make to
turbulence in ionized gas? {\it Mac Low:} In our models, most of the
turbulence comes from the interaction of multiple shells.  Diffuse
ionizing radiation will ionize some of that gas, producing diffuse,
turbulent, ionized gas.  H~{\sc ii} regions also contribute, of
course. In Mac Low \& Klessen \cite{mk04} we use results from Matzner
\cite{cdm02} to argue that H~{\sc ii} region expansion is only a minor
(< 1\%) contributor to ISM kinetic energy.\\

\noindent
{\it Heiles:} You emphasized the breadth of the pressure
distribution.  But it's really not more than an order of magnitude, right?
{\it Mac Low:} That is true for the volume-weighted FWHM.  However, a
mass-weighted view shows that a substantial fraction of the mass is at
the high-density end.

\end{document}